\documentclass[traditabstract,longauth]{aa}
\usepackage{graphicx} 
\usepackage{txfonts}
\usepackage{natbib}
\bibpunct{(}{)}{;}{a}{}{,}

\begin{document}

\title{Far-Infrared Line Imaging of the Starburst Ring in \object{NGC~1097} with the \textit{Herschel}/PACS Spectrometer\thanks{\textit{Herschel} is an ESA space observatory with science instruments provided by European-led Principal Investigator consortia and with important participation from NASA.}}
\titlerunning{\textit{Herschel}/PACS Spectroscopy of \object{NGC~1097}}
   \author{P.\ Beir\~ao\inst{1}
          \and L.\ Armus\inst{1}
           \and P. N.\ Appleton\inst{2}
         \and J.-D. T.\ Smith\inst{3}
           \and K. V.\ Croxall\inst{3}
       \and E. J.\ Murphy\inst{1}
        \and D. A.\ Dale\inst{4}
           \and G.\ Helou\inst{2}
           \and R. C.\ Kennicutt\inst{5}
           \and D.\ Calzetti\inst{6}
         \and A. D.\ Bolatto\inst{7}
        \and B. R.\ Brandl\inst{8}
           \and A. F.\ Crocker\inst{6}
         \and B. T.\ Draine\inst{9}
          \and G.\ Dumas\inst{10}   
          \and C. W.\ Engelbracht\inst{11}
          \and A.\ Gil de Paz\inst{12}
          \and K. D.\ Gordon\inst{13}
           \and B.\ Groves\inst{8}
          \and C.-N.\ Hao\inst{14}
         \and J. L.\ Hinz\inst{11}
           \and L. K.\ Hunt\inst{15}
          \and B. D.\ Johnson\inst{5}
           \and J.\ Koda\inst{16}
           \and O.\ Krause\inst{10}
           \and A. K.\ Leroy\inst{17}\thanks{Hubble Fellow}
          \and S. E.\ Meidt\inst{10}
          \and J.\ Richer\inst{18}
           \and H.-W.\ Rix\inst{10}
         \and N.\ Rahman\inst{19}
           \and H.\ Roussel\inst{21}
          \and K. M.\ Sandstrom\inst{10}
           \and M.\ Sauvage\inst{22}
           \and E. Schinnerer\inst{10}
          \and R. A.\ Skibba\inst{11}
          \and S.\ Srinivasan\inst{21}
           \and F.\ Walter\inst{10}
           \and B. E.\ Warren\inst{19}
           \and C. D.\ Wilson\inst{20}
           \and M. G.\ Wolfire\inst{7}
          \and S.\ Zibetti\inst{10}
      }

   \institute{\textit{Spitzer} Science Center, California Institute of Technology, MC
              314-6, Pasadena, CA 91125, USA \\
          \email{pedro@ipac.caltech.edu}
                   \and NASA \textit{Herschel} Science Center, IPAC, California Institute of
              Technology, Pasadena, CA 91125, USA
           \and Department of Physics and Astronomy, Mail Drop 111, University
              of Toledo, 2801 West Bancroft Street, Toledo, OH 43606, USA
                 \and Department of Physics \& Astronomy, University of Wyoming,
              Laramie, WY 82071, USA           
         \and Institute of Astronomy, University of Cambridge, Madingley Road,
              Cambridge CB3 0HA, UK
         \and Department of Astronomy, University of Massachusetts, Amherst,
              MA 01003, USA
             \and Department of Astronomy, University of Maryland, College Park,
              MD 20742, USA
          \and Leiden Observatory, Leiden University, P.O. Box 9513, 2300 RA
              Leiden, The Netherlands
         \and Department of Astrophysical Sciences, Princeton University,
              Princeton, NJ 08544, USA
            \and Max-Planck-Institut f\"{u}r Astronomie, K\"{o}nigstuhl 17, 69117
              Heidelberg, Germany
     \and Steward Observatory, University of Arizona, Tucson, AZ 85721,
              USA 
          \and Departamento de Astrofisica, Facultad de Ciencias Fisicas,
              Universidad Complutense Madrid, Ciudad Universitaria, Madrid, E-28040, Spain
        \and Space Telescope Science Institute, 3700 San Martin Drive,
              Baltimore, MD 21218, USA
           \and Tianjin Astrophysics Center, Tianjin Normal University, 
              Tianjin 300387, China
     \and INAF - Osservatorio Astrofisico di Arcetri, Largo E. Fermi 5,
              50125 Firenze, Italy
     \and Department of Physics and Astronomy, SUNY Stony Brook, Stony
              Brook, NY 11794-3800, USA
          \and National Radio Astronomy Observatory, 520 Edgemont Road, 
              Charlottesville, VA 22903, USA
        \and Astrophysics Group, Cavendish Laboratory, J. J. Thomson Avenue,
              Cambridge CB3 0HE; Kavli Institute for Cosmology, c/o Institute
              of Astronomy, University of Cambridge, Madingley Road, Cambridge
              CB3 0HA
        \and ICRAR, M468, University of Western Australia, 35 Stirling Hwy, Crawley, WA, 6009, Australia 
        \and Deptartment of Physics \& Astronomy, McMaster University, Hamilton,
              Ontario L8S 4M1, Canada
    \and Institut d'Astrophysique de Paris, UMR7095 CNRS
              Universit\'e Pierre \& Marie Curie, 98 bis boulevard Arago, 75014 Paris, France
        \and CEA/DSM/DAPNIA/Service d'Astrophysique, UMR AIM, CE Saclay,
              91191 Gif sur Yvette Cedex
         }
\abstract
{\object{NGC~1097} is a nearby SBb galaxy with a Seyfert nucleus and a bright starburst ring.
We study the physical properties of the interstellar medium (ISM) in the ring using spatially resolved far-infrared spectral maps of the circumnuclear starburst ring of \object{NGC~1097}, obtained with the PACS spectrometer on board the \textit{Herschel} Space Telescope. In particular, we map the important ISM cooling and diagnostic emission lines of [OI] 63 $\mu$m, [OIII] 88 $\mu$m, [NII] 122 $\mu$m, [CII] 158 $\mu$m and [NII] 205 $\mu$m. We observe that in the [OI] 63 $\mu$m, [OIII] 88 $\mu$m, and [NII] 122 $\mu$m line maps, the emission is enhanced in clumps along the NE part of the ring. We observe evidence of rapid rotation in the circumnuclear ring, with a rotation velocity of $\sim220$ km s$^{-1}$ (inclination uncorrected) measured in all lines.
The [OI] 63 $\mu$m/[CII] 158 $\mu$m ratio varies smoothly throughout the central region, and is enhanced on the northeastern part of the ring, which may indicate a stronger radiation field. This enhancement coincides with peaks in the [OI] 63 $\mu$m and [OIII] 88 $\mu$m maps. Variations of the [NII] 122 $\mu$m/[NII] 205 $\mu$m ratio correspond to a range in the ionized gas density between 150 and 400 cm$^{-3}$.}

\maketitle

\section{Introduction}
\label{intro}


\object{NGC~1097} is a Seyfert 1 galaxy with a bright starburst ring with a diameter of 2 kpc and a strong large-scale bar \citep{gerin88,kohno03,hsieh08} with a length of 15 kpc. Optical and near-infrared images reveal dust lanes that run along the primary large-scale (15 kpc) bar and curve into the ring, which is formed by two very
tight spiral arms, and a second bar inside the ring \citep{quillen95,prieto05}. This bar may be responsible for driving gas into the nucleus, possibly fueling the central super-massive black hole \citep{prieto05,fathi06,davies09}, and may also have triggered the formation
of a compact star cluster seen near the nucleus. \object{NGC~1097} provides an excellent opportunity to study the physical conditions of the interstellar medium (ISM) in a nearby galaxy with both a starburst and an active nucleus.

The nucleus and the star-forming ring are prominent in CO and HCN line emission \citep{kohno03}. 
Near-infrared spectroscopy \citep{reunanen02, kotilainen00} reveals emission from both ro-vibrational H$_2$ and H-recombination lines at the nucleus, the star-forming ring, and the region in between. Optical long-slit spectroscopy \citep{storchi96} shows strong ionized gas emission from both the nucleus and the ring, along with faint line emission from the inner region, exhibiting a LINER-type spectrum. 

With \textit{Herschel}/PACS \citep{poglitsch10} we are now able to target the most important cooling lines of the warm ISM on physical scales much smaller than ever before possible in external galaxies. The  KINGFISH project (Key Insights on Nearby Galaxies: a Far- Infrared Survey with \textit{Herschel} - PI: R. C. Kennicutt) is an open-time \textit{Herschel} key program which aims to measure the heating and cooling of the gaseous and dust components of the ISM in a sample of 61 nearby galaxies with the PACS and SPIRE instruments. The far infrared spectral range covered by PACS includes several of the most important cooling lines of the atomic gas, notably [CII] 158 $\mu$m, [OI] 63 $\mu$m, [OIII] 88 $\mu$m, [NII] 122 $\mu$m, and [NII] 205 $\mu$m. NGC 1097 is one of the KINGFISH targets selected for the \textit{Herschel} Science Demonstration Program (SDP) (for PACS imaging of NGC 1097 see \citet{sandstrom10} in this volume and for SPIRE observations see \citet{engel10} in this volume).

In this letter, we present far-infrared spectral line maps of the circumnuclear starburst ring and the large-scale bar of \object{NGC~1097}, obtained with the PACS spectrometer on board the ESA \textit{Herschel} Space Telescope \citep{pilbratt10}. The maps presented in this letter are the first PACS spectral maps from the KINGFISH program. Throughout this paper we assume a distance to \object{NGC~1097} of $19.1$ Mpc \citep{willick97}, which gives a projected scale of $1\arcsec=92$ pc.

\section{Observations and data reduction}


\begin{figure}
\centering
\includegraphics[width=0.4\textwidth]{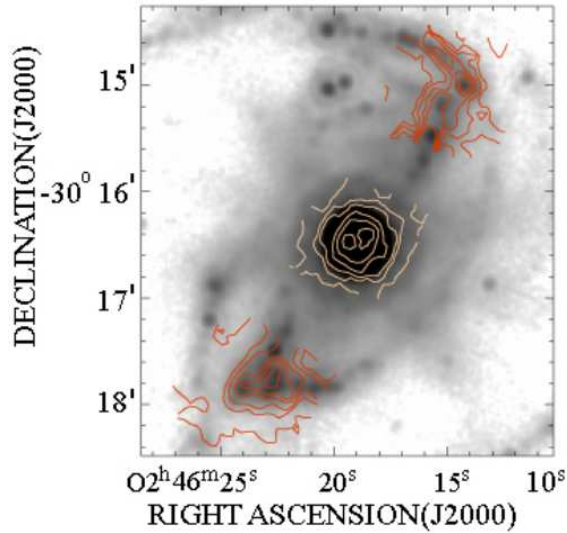}
\caption{Composite of [CII] 158 $\mu$m emission maps in the center (yellow contours) and extranuclear positions (red contours), overlaid on a \textit{Spitzer}/MIPS $24\mu$m image of \object{NGC~1097} (in grayscale). The extranuclear maps are a combination of wavelength-switching data (North) and chop-nod data (South). The contour levels in units of 10$^{-7}$ W/m$^2$/sr are: 0.35, 0.70, 1.4, 2.8, 4.2, and 5.95 for the nucleus, 0.092, 0.184, 0.276, 0.368, 0.460, and 0.570 for the southern extranuclear region, and 0.051, 0.102, 0.153, 0.204, and 0.238 for the northern extranuclear region.}
\label{ciicolor}
\end{figure}

\object{NGC~1097} was observed with the $5\times5$ pixel integral field unit (IFU) of the PACS Spectrometer in both chop-nod (CN) and wavelength
switching (WS) modes as part of a series of Science Demonstration Phase tests. The field of view of the IFU is 
$47\arcsec\times47\arcsec$ with $9\farcs4$ pixels \citep{poglitsch10}. The diffraction limited FWHM beam size of PACS is $5\farcs2$ at $60 - 85\mu$m and $12\arcsec$ at $130 - 210\mu$m. The spectral resolution of PACS is 180 km\,s$^{-1}$ for [OI] 63 $\mu$m, 120 km\,s$^{-1}$ for [OIII] 88 $\mu$m, 290 km\,s$^{-1}$ for [NII] 122 $\mu$m, 240 km\,s$^{-1}$ for [CII] 158 $\mu$m, and 150 km\,s$^{-1}$ for [NII] 205 $\mu$m.
CN observations were carried out between January 4 and January 22, 2010, with an observing time of 19.4 hrs, with 17.7 hrs on-source. WS observations were carried out between January 4 and January 22, 2010, and the observing time was 6.6 hrs, of which 3.6 hrs are on-source. Five different regions were observed: the nucleus, two extranuclear positions at the outer ends of the bar, and one strip at each side of the nucleus, following the minor axis of the galaxy. 
CN observations were obtained in a $2\times2$ raster in the nucleus for all five target lines (with a step size of $9\farcs4$), in a single short strip along the minor axis ([OI] 63 $\mu$m, [NII] 122 $\mu$m, and [CII] 158 $\mu$m), and in a single extranuclear position to the south ([OI] 63 $\mu$m, [OIII] 88 $\mu$m, [NII] 122 $\mu$m, and [CII] 158 $\mu$m). Each map covers an area of $4.5\times4.5$ kpc$^2$. 
In the basic CN mode, the telescope chops between two regions of sky $6\arcmin$ apart, with the target placed
alternately in one or the other beam. WS observations were obtained for the same lines in a $2\times2$ raster in the nucleus, two short strips along the minor axis on each side of the nucleus, and in two extranuclear positions to the north and south of the nucleus.
Line fluxes obtained from WS data in the same regions were in all cases comparable within the flux uncertainties to CN results, and the KINGFISH survey will utilize WS for the remainder of the sample. Here we focus on the CN observations, which have the best signal to noise ratios.


The spectra were reduced using HIPE (\textit{Herschel} Interactive Processing Environment), Special Development version 3.0 CIB 1134. The pipelines remove detector artifacts, obvious cosmic ray signatures and apply ground-based flat-field corrections. Dark-current subtraction was done
for the WS case. Additional band-dependent scale factors were applied to the post-processed extracted fluxes to adjust the ground-based
absolute calibration to in-flight observations of calibrators\footnote{Spectral extractions followed the prescription described in the \textit{Herschel} PACS Data Reduction Guide
and post-launch corrections to the ``nominal response" followed the guidelines of Version 1.0 of the Spectroscopy and Calibration PACS release note.}. The uncertainty in the line fluxes is dominated by the absolute flux calibration. The flux calibration uncertainties are on the order of 30\%, and pixel-to-pixel relative calibration uncertainties are on the order of $10\%$.

\section{Results}

\begin{figure}
\centering
\includegraphics[width=0.5\textwidth]{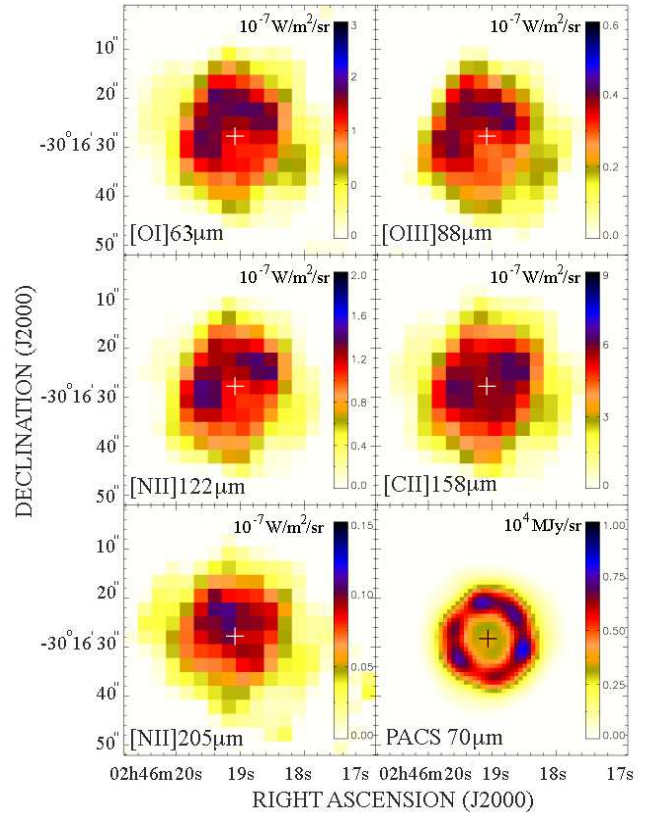}
\caption{Spectral maps of the lines observed with PACS at the nuclear position, and an image of the ring at 70 $\mu$m with PACS. The maps represent the integrated flux over $-500<$v$_{sys}< 500$ km\,s$^{-1}$, with v$_{sys}=1271$ km s$^{-1}$. North is up and East is to the left. In all cases the size of the image is $51\arcsec\times51\arcsec$. The cross marks the location of the radio position of the nucleus.}
\label{maps}
\end{figure}




In Fig. 1 we present an overlay of the [CII] 158 $\mu$m maps of the nuclear and extranuclear regions of NGC 1097 on a 24 $\mu$m image taken with the Multiband Imaging Photometer for \textit{Spitzer} MIPS \citep{rieke04}. The $24 \mu$m emission is well traced by the [CII] 158 $\mu$m emission in all the regions where [CII] 158 $\mu$m emission is observed. As there are no CN observations of the northern extranuclear region, we show a WS map, which clearly follows the detailed structure evident in the $24\mu$m image. The signal-to-noise ratio in the minor axis strips is very low, and these data are not shown in Fig. 1.

In Fig. 2 we present continuum-subtracted maps of the center of \object{NGC~1097}, for the [CII] 158 $\mu$m, [OI] 63 $\mu$m, [OIII] 88 $\mu$m, [NII] 122 $\mu$m, and [NII] 205 $\mu$m emission lines. For comparison, we also show a PACS 70$\mu$m continuum image (see \citet{sandstrom10} in this issue). 
The [OI] 63 $\mu$m, [OIII] 88 $\mu$m, and [NII] 122 $\mu$m maps are the most similar. There is a partial clumpy ring-like structure, with peaks NW, N and E of the nucleus in the [OI] 63 $\mu$m and [OIII] 88 $\mu$m maps, whereas the N clump is absent in the [NII] 122 $\mu$m map. The peaks observed in the [CII] 158 $\mu$m coincide with the clumps observed in the [NII] 122 $\mu$m map, but the overall distribution is much smoother. Only the N and NW peaks have a counterpart in the PACS $70\mu$m image.  
The [NII] 205 $\mu$m looks strikingly different than the other maps. It has a peak NE of the nucleus but lacks the NW and E hotspots present in the other maps. The lack of a resolved ring in the [CII] 158 $\mu$m and [NII] 205 $\mu$m may be partially due to the large beam at these wavelengths. A careful measurement of the spectrometer PSF on $10\arcsec$ scales will be required in order to further separate real physical components in a system as small as the \object{NGC~1097} ring. 


\begin{figure}
\centering
\includegraphics[width=0.4\textwidth]{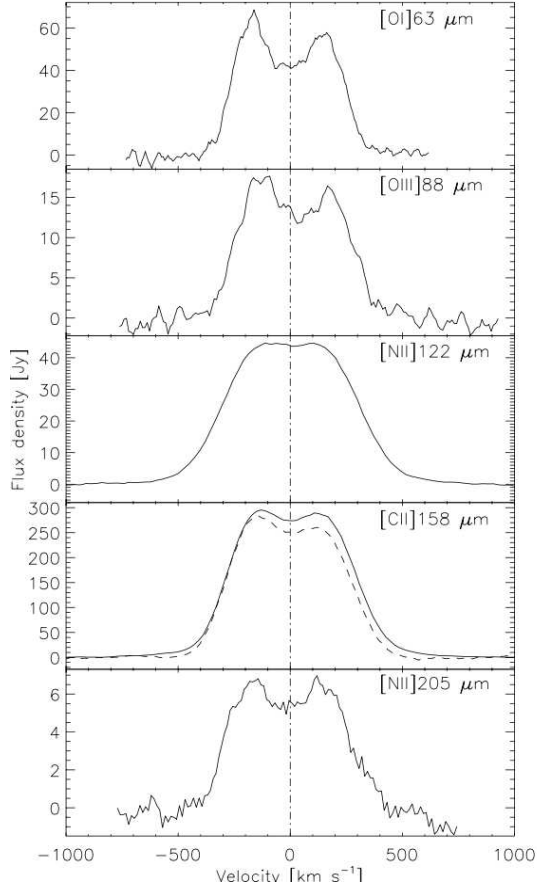}
\caption{Integrated spectra of the nucleus and ring. The system velocity v$_{sys}=1271$ km\,s$^{-1}$ is set as the zero-point, marked by the dashed-dotted line. The dashed line is the spectrum of the [CII] emission in wavelength-switching mode.}
\end{figure}

To extract integrated spectra from the line maps, we selected a circular region of $23\farcs8$ in radius centered on the nucleus which includes the ring, but avoids the noisy edges of the map. 
The continuum subtracted velocity profiles for each line are shown in Fig. 3. In all cases, zero velocity was defined as v$_{sys}=1271$ km\,s$^{-1}$.
An overplot of the [CII] 158 $\mu$m velocity profile made from the WS data is also shown for comparison, which has a peak flux of $\sim90$\% of the CN peak, within the flux uncertainty. The smooth appearance of the [CII] 158 $\mu$m and the [NII] 122 $\mu$m line profiles is due to the worse spectral resolution at these wavelengths.
The small difference is attributed to transients in the CN data which are not present in the WS data. The most notable characteristic of these profiles is that they are all double-peaked, which is expected for a rotating ring. 
In Fig. 4 we show the velocity map of [OI] 63 $\mu$m.
The velocity spans a range of $\pm220$ km\,s$^{-1}$, being redshifted to the SE, and blueshifted to the NW. The [OI] velocity field is consistent with circular rotation also seen in ionized (H$\alpha$, \citet{fathi06}) and molecular gas (CO (2-1) \citet{hsieh08}). In Table 1 we present the integrated line fluxes for the nucleus and extranuclear positions. 

\begin{figure}
\centering
\includegraphics[width=0.4\textwidth]{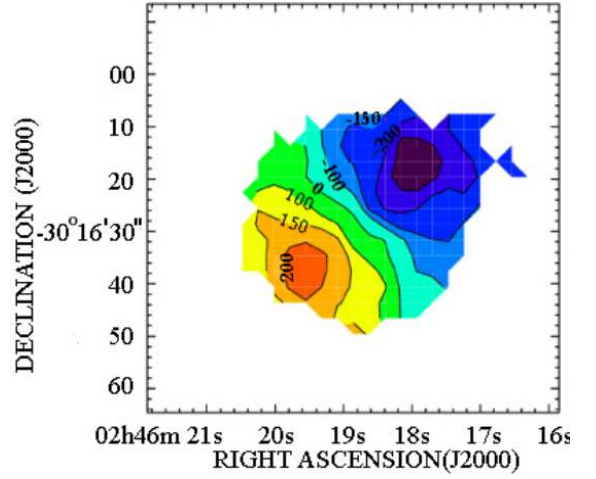}
\caption{Velocity field of the [OI]  line in the center of \object{NGC~1097}. The contours represent approaching (blue) and receding (red) emission relative to the systemic velocity, v$_{sys}=1271$ km\,s$^{-1}$.}
\end{figure}

   \begin{table}
   \centering
      \caption{Integrated line fluxes$^{1}$ ($10^{-15}$ Wm$^{-2}$)}
         \begin{tabular}{cccccc}
            \hline
            \hline
            \noalign{\smallskip}
Position  	& [OI] & [OIII] & [NII] 	& [CII] & [NII]	\\
 & 63 $\mu$m & 88$\mu$m & 122$\mu$m & 158$\mu$m & 205$\mu$m\\
            \noalign{\smallskip}
            \hline
            \noalign{\smallskip}
Nucleus & 3.5 & 0.7	& 2.0 & 9.0	& 0.15   \\ 
Enuc S & 0.28 & 0.13 & 0.07 & 1.2 & - \\
Enuc N & - & - & - & 0.9 & - \\
         \noalign{\smallskip}
            \hline
         \end{tabular}
         \footnotetext{}{$^{1}$``Nucleus'' comprises a circular region of $23\farcs8$ in radius ($\sim 2$ kpc) centered on the Seyfert nucleus. The fluxes for the other positions were measured over the whole maps ($50\arcsec\times50\arcsec \sim4\times4$ kpc). 
         All measured fluxes have an uncertainty of 30\%.}
          \end{table}

\section{Diagnosing the ionized and neutral gas properties}


With the line maps we can study spatial variations in the properties of the warm atomic gas. Fine structure line ratios trace the intensity of the incident radiation field $G_0$ on the neutral gas and the electron gas density $n_e$, respectively \citep[e.g.,][]{kaufman99}.
In Fig. 5 we show maps of the [OI] 63 $\mu$m/[CII] 158 $\mu$m and the [NII] 122 $\mu$m/[NII] 205 $\mu$m line ratios, overlaid with the contours of the PACS $70\mu$m image. Before making the ratio maps, each image was smoothed to the beam size at the longer wavelength using a Gaussian profile, while conserving flux. The [OI] 63 $\mu$m/[CII] 158 $\mu$m ratio varies smoothly with values ranging between $0.25-0.45$ throughout the central region, and is enhanced on the northeast part of the ring. This enhancement partially coincides with the peaks in the [OI] 63 $\mu$m and [OIII] 88 $\mu$m line maps and the peak of the H$\alpha$ emission \citep{hummel87}, which indicates that the [OI] 63 $\mu$m/[CII] 158 $\mu$m ratio traces the most massive star forming knots in the ring. The values of [OI] 63 $\mu$m/[CII] 158 $\mu$m are a factor of 2 lower than for the starburst galaxy \object{M82} \citep{colbert99}, but similar to the values found for most star-forming galaxies \citep{malhotra01}. The peak in the [OI] 63 $\mu$m/[CII] 158 $\mu$m map coincides with the N peak in the PACS 70$\mu$m map, suggesting that at this location, a star formation rate density higher than in the rest of the ring is enhancing both dust emission and [OI] 63 $\mu$m line emission. 
The ratio is also enhanced on a region where the ring and the dusty spiral arms intersect, which indicate a possible shock contribution to the [OI] 63 $\mu$m flux. Models of emission lines from shocks \citep{holl89} predict values of the [OI] 63 $\mu$m/[CII] 158 $\mu$m ratio of at least 10. Our measured value of $\sim0.4$ are at odds with a pure-shock interpretation of the line ratios. On the other hand, line emission from shocks cannot be ruled out because the region in the PACS beam could be a composite of both shocked gas and additional [CII] flux contributed from star formation.
Models of line emission from photodissociation regions (PDRs) around O stars \citep[e.g.,][]{kaufman99} show that [OI] 63 $\mu$m/[CII] 158 $\mu$m increases with the radiation field intensity $G_0$ and neutral gas density $n$. However, at low densities, [OI] 63 $\mu$m/[CII] 158 $\mu$m is mostly sensitive to $G_0$. Using \citet{kaufman99} models for $n<10^4$ cm$^{-3}$, we estimate $G_0$ to lie between $60-300$, and using \citet{meijerink07}, $G_0$ is estimated between $200-600$. The higher $G_0$ found using \citet{meijerink07} is likely due to their lower adopted O/C abundance ratio.

The [NII] 122 $\mu$m/[NII] 205 $\mu$m ratio varies between $4.0-6.0$ throughout the central $40\arcsec$. The values that correspond to the [NII] 122 $\mu$m peaks are 4.8 and 4.5, whereas at the [NII] 205 $\mu$m peak, the [NII] 122 $\mu$m/[NII] 205 $\mu$m ratio is 4.0. From the comparison with the $70\mu$m contours, we see that the highest values of the [NII] 122 $\mu$m/[NII] 205 $\mu$m ratio coincide with the ring, while the lowest are found in the inner region of the ring and also where the dust lanes meet the ring.
The variations in the [NII] 122 $\mu$m/[NII] 205 $\mu$m ratio are therefore due to a variation in the ionized gas density between the ring and the inner region. 
Using a five level model of N$^{+}$ we find that the variation of [NII] 122 $\mu$m/[NII] 205 $\mu$m ratios between the inner region and the ring corresponds to a variation of the electron density from 150 to 400 cm$^{-3}$. The results are insensitive to typical gas temperatures ($T\sim6000-10000$ K) in photoionized gas. This means that the density increases by at least a factor of 5 in the ring compared to the inner region and the region where the dust lane and the ring meet. These values are similar to the central region of \object{M82}, in which a mean ratio value of $4.2^{+1.6}_{-1.2}$ was measured, corresponding to a mean electron density of $180^{+209}_{-120}$ cm $^{-3}$ across the central $50\arcsec$ \citep{petuch94}. This is also consistent with the value of $\sim220$ cm$^{-3}$ from the mid-infrared [SIII]18$\mu$m/33$\mu$m ratio over the same region \citep{dale07}.

The peak in the [NII] 205 $\mu$m emission line map in Fig. 2 has a line flux of $\sim 0.11\times10^{-7}$ W\,m$^{-2}$\,sr$^{-1}$ and a [CII] 158 $\mu$m/[NII] 205 $\mu$m $\sim 45$. For $150 < n_{\mathrm{e} }< 400$ cm$^{-3}$ the [CII] 158 $\mu$m/[NII] 205 $\mu$m ratio in ionized gas is expected to be $\sim3$ \citep{oberst06} and thus most ($\sim 90 \%$) of the [CII] 158 $\mu$m emission we measure is coming from neutral gas. 
The gas heating efficiency, measured by the ([OI] + [CII])/FIR ratio, seems to stay constant in the mapped area. The log of the ratio is $\sim-2.2$ inside the ring and $\sim-2.3$ on the ring, both consistent with the values found in nearby galaxies by \citet{malhotra01}.

In summary, we have used the PACS Spectrometer to map the [OI] 63 $\mu$m, [OIII] 88 $\mu$m, [CII] 158 $\mu$m, [NII] 122 $\mu$m and 206$\mu$m far-infrared cooling line emission in the central 5 kpc of \object{NGC~1097} for the first time.  While the [OI] 63 $\mu$m, [OIII] 88 $\mu$m and [NII] 122 $\mu$m line maps appear qualitatively similar, the [NII] 205 $\mu$m map shows a different distribution. The [OI] 63 $\mu$m/[CII] 158 $\mu$m map shows a relative hotspot on the NE portion of the ring indicative of a stronger radiation field or a region of shocked gas.  The [NII] 122 $\mu$m/[NII] 205 $\mu$m map shows a clear increase of ionized gas density in the ring, associated with massive star formation activity.
      
\begin{figure}
\centering
\includegraphics[width=0.45\textwidth]{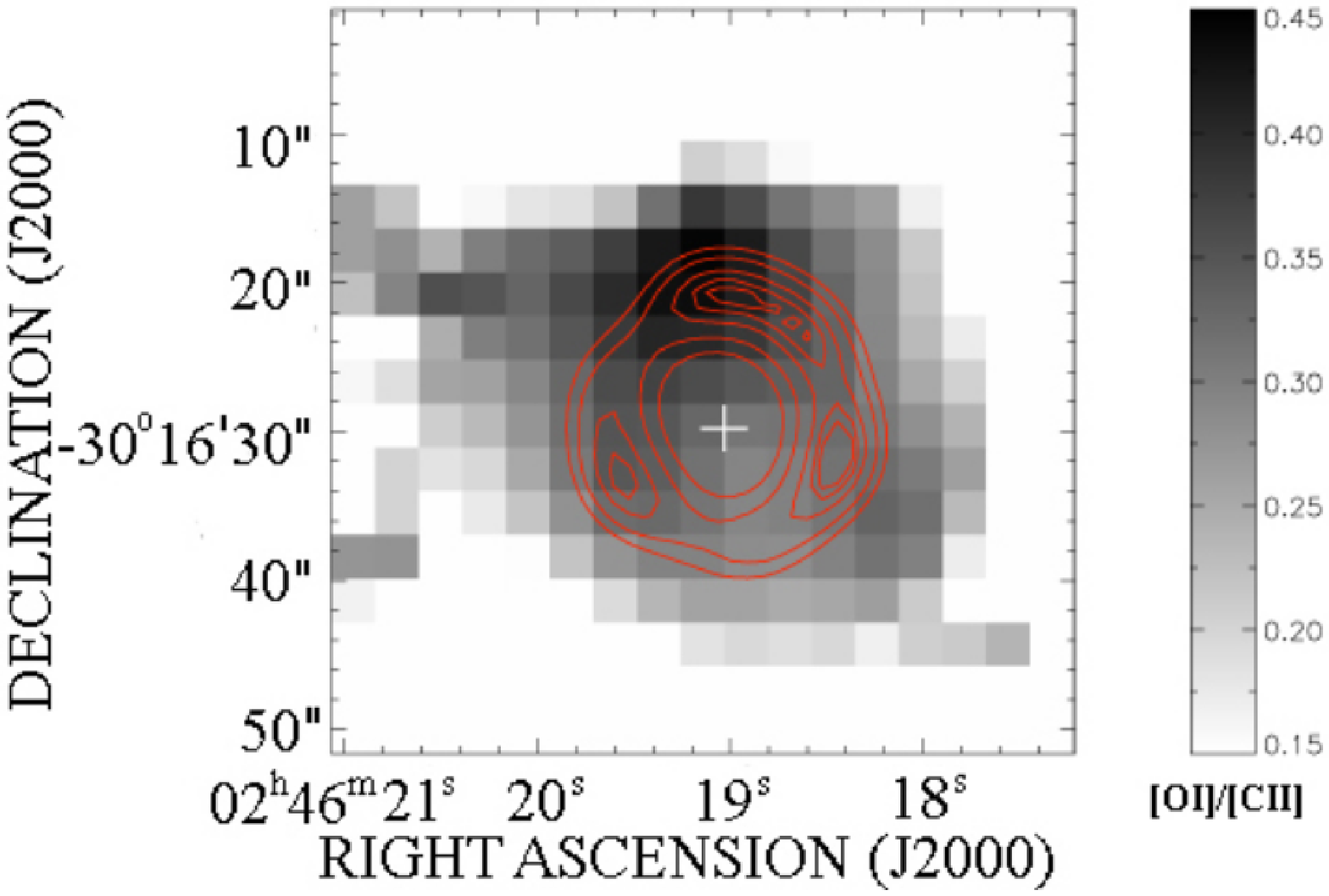}
\includegraphics[width=0.45\textwidth]{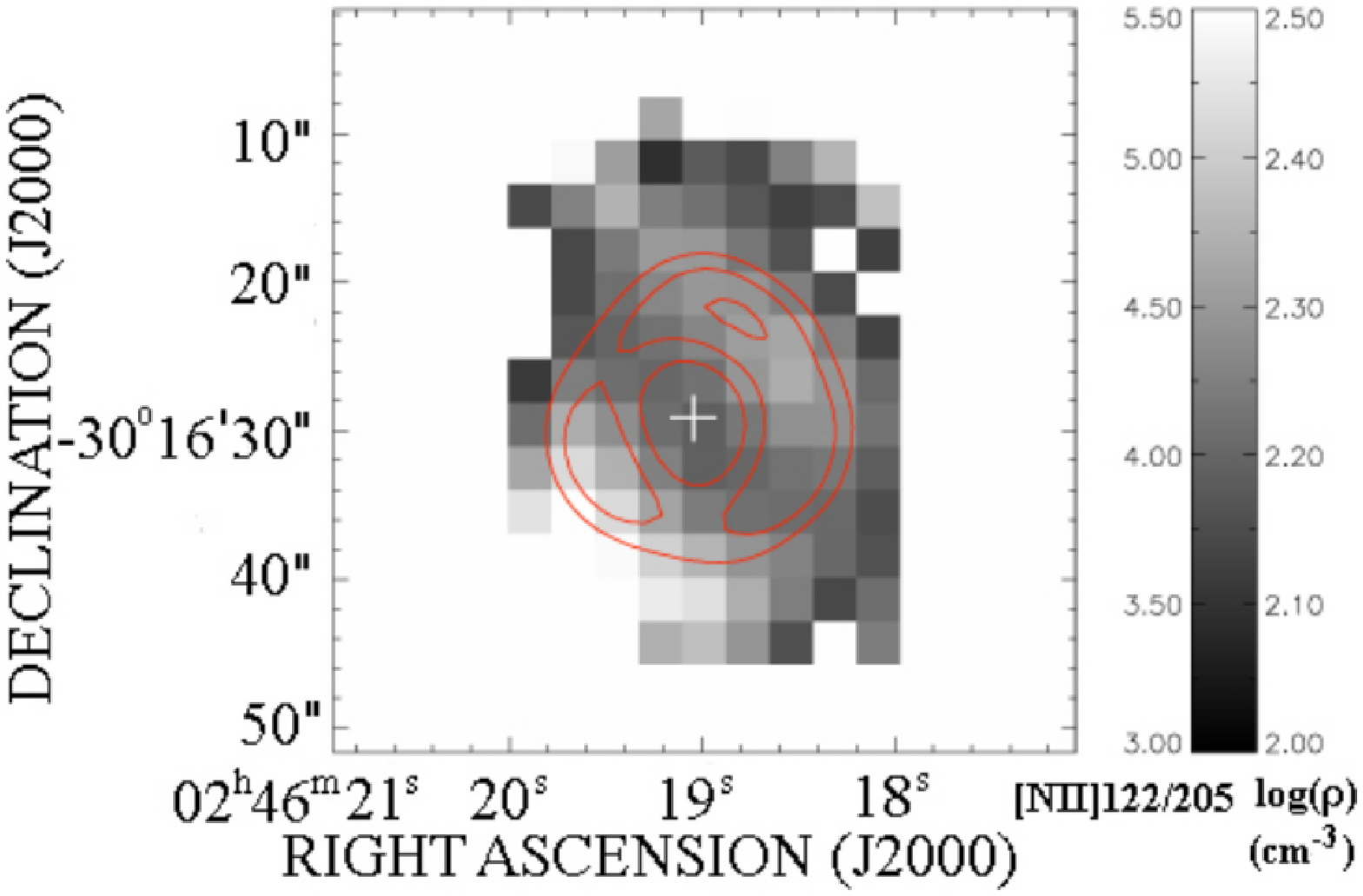}
\caption{Top: map of the [OI] 63 $\mu$m/[CII] 158 $\mu$m ratio. The [OI] 63 $\mu$m map was smoothed to match the resolution at $158\mu$m. Bottom: map of the [NII] 122 $\mu$m/[NII] 205 $\mu$m ratio. The [NII] 122 $\mu$m map was smoothed to match the [NII] 205 $\mu$m map. Both maps are overlaid with contours from the PACS $70\mu$m map, after smoothing to the resolution of each map. All the ratio maps were built from line maps clipped at a $2\sigma$ level above the noise. The cross marks the location of the nucleus.
}
\label{spectra}
\end{figure}


\begin{acknowledgements}

This work is based [in part] on observations made with \textit{Herschel}, a European Space Agency Cornerstone Mission with significant participation by NASA. Support for this work was provided by NASA through an award issued by JPL/Caltech. PACS has been developed by a consortium of institutes led by MPE (Germany) and including UVIE (Austria); KUL, CSL, IMEC (Belgium); CEA, OAMP (France); MPIA (Germany); IFSI, OAP/AOT, OAA/CAISMI, LENS, SISSA (Italy); IAC (Spain). This development has been supported by the funding agencies BMVIT (Austria), ESA-PRODEX (Belgium), CEA/CNES (France), DLR (Germany), ASI (Italy), and CICT/MCT (Spain). Data presented in this paper were analyzed using ÒThe \textit{Herschel} Interactive Processing Environment (HIPE),Ó a joint development by the \textit{Herschel} Science Ground Segment Consortium, consisting of ESA, the NASA \textit{Herschel} Science Center, and the HIFI, PACS and SPIRE consortia. We would like to thank Dario Fadda and Jeff Jacobson for software support and Hanae Inami for the help on the velocity diagram. 

\end{acknowledgements}

\end{document}